\documentclass[aps,pra,preprint,groupedaddress]{revtex4-1}

\usepackage{graphicx}

\begin{document}
	

\title{Matter-wave interferometry with atoms in high Rydberg states}


\author{J. E. Palmer}
\author{S. D. Hogan}

\affiliation{Department of Physics and Astronomy, University College London, Gower Street, London WC1E 6BT, United Kingdom}
	

\date{\today}

\begin{abstract}
	Matter-wave interferometry has been performed with helium atoms in high Rydberg states. In the experiments the atoms were prepared in coherent superpositions of Rydberg states with different electric dipole moments. Upon the application of an inhomogeneous electric field, the different forces on these internal state components resulted in the generation of coherent superpositions of momentum states. Using a sequence of microwave and electric field gradient pulses the internal Rydberg states were entangled with the momentum states associated with the external motion of these matter waves. Under these conditions matter-wave interference was observed by monitoring the populations of the Rydberg states as the magnitudes and durations of the pulsed electric field gradients were adjusted. The results of the experiments have been compared to, and are in excellent quantitative agreement with, matter-wave interference patterns calculated for the corresponding pulse sequences. For the Rydberg states used, the spatial extent of the Rydberg electron wavefunction was $\sim320$~nm. Matter-wave interferometry with such giant atoms is of interest in the exploration of the boundary between quantum and classical mechanics. The results presented also open new possibilities for measurements of the acceleration of Rydberg positronium or antihydrogen atoms in the Earth's gravitational field. \bigskip
\end{abstract}

\maketitle

\section{Introduction}\label{sec:introduction}

The use of inhomogeneous electric fields to control the motion of atoms or molecules in Rydberg states with high principal quantum number, $n$, was first suggested by Wing~\cite{wing80a}, and Breeden and Metcalf~\cite{breeden81a} in the early 1980s. However, it was not until work by Softley and co-workers in the late 1990s that experiments were performed in which such control was demonstrated~\cite{townsend01a}. In these experiments, krypton Rydberg atoms travelling in pulsed supersonic beams were state-selectively deflected in an inhomogeneous dipolar electric field with a gradient perpendicular to the direction of propagation of the atoms. This apparatus represented an electric analogue of that of Gerlach and Stern in which space quantisation was first identified~\cite{gerlach22a,gerlach22b}. These experiments were followed by further deflection, and ultimately acceleration and deceleration experiments with beams of H$_2$~\cite{yamakita04a}. This early work subsequently led to the development of the methods of Rydberg-Stark deceleration, initiated by Merkt and co-workers, for controlling the translational motion of, and trapping, atoms and molecules in high Rydberg states. The experimental tools that have been developed in this context include deflectors~\cite{townsend01a,yamakita04a,allmendinger14a}, guides~\cite{lancuba13a,ko14a,deller16a,deller19a}, velocity selectors~\cite{alonso17a}, lenses~\cite{vliegen06a}, mirrors~\cite{vliegen06b,jones17a}, beamsplitters~\cite{palmer17a}, decelerators~\cite{yamakita04a,vliegen04a,vliegen05a}, and traps~\cite{vliegen07a,hogan08a,hogan09a,seiler11a,hogan12a,lancuba16a}. These have been implemented with atoms composed of matter and antimatter, and with molecules. The methods of Rydberg-Stark deceleration have contributed to new approaches to the study of low-temperature ion-molecule reactions~\cite{allmendinger16a,allmendinger16b}, allowed investigations of excited-state decay processes, including effects of blackbody radiation, on time scales that were not previously possible~\cite{seiler16a}, and played an essential role in recent developments in positronium physics~\cite{cassidy18a}.

In all Rydberg-Stark deceleration experiments up to now, the atoms or molecules were prepared in selected Rydberg states by laser photoexcitation, and, under the conditions in which the experiments were performed, quantisation of the motional states of the samples could be neglected. Consequently, this set of methodologies can be classified as incoherent Rydberg atom or molecule optics. Here we present the results of Rydberg-Stark deceleration experiments performed with atoms prepared in coherent superpositions of Rydberg states with different electric dipole moments. In the presence of inhomogeneous electric fields the different forces on these internal state components allow the generation of superpositions of momentum states which have been exploited for Rydberg-atom interferometry. Experiments of the kind reported here may be classified as coherent Rydberg atom optics. The results presented open new opportunities in the exploration of the boundary between quantum and classical mechanics, the study of spatial decoherence in large quantum systems~\cite{bassi13a}, and measurements of the acceleration of neutral particles composed of antimatter, e.g., Rydberg positronium or antihydrogen, in the gravitational field of the Earth~\cite{mills02a,kellerbauer08a,amole13a,cassidy14a}.

Matter-wave interferometry~\cite{cronin09a,hornberger12a} in which the forces exerted by inhomogeneous electric or magnetic fields on atoms or molecules in coherent superpositions of internal states with different dipole moments was considered from a theoretical perspective since the early 1950s~\cite{bohm51a,wigner63a,schwinger88a}. This led to the realisation of an early electric atom interferometer~\cite{sokolov70a,sokolov73a}, and later the magnetic longitudinal Stern-Gerlach atom interferometer~\cite{miniatura91a,miniatura92a,nicChormaic93a} -- both demonstrated with hydrogen atoms in metastable $n=2$ levels. In the latter, atoms in coherent superpositions of Zeeman sublevels travelled through a static inhomogeneous magnetic field. As they entered the field, acceleration and deceleration of the two components of the initial internal state superposition resulted in the generation of a superposition of momentum states each with a different de Broglie wavelength, $\lambda_{\mathrm{dB}}$. When the atoms exited the field the subsequent deceleration and acceleration of the two matter-wave components returned their momenta to the initial value. However, because of their different flight times through the magnetic field a spatial separation between the centres of mass of the two matter waves remained. This spatial separation, if equal to $\lambda_{\mathrm{dB}}$ ($\lambda_{\mathrm{dB}}/2$) led to constructive (destructive) matter-wave interference which was monitored by projecting the final superposition of Zeeman sublevels onto the stationary basis states. Interference fringes were observed by monitoring the internal quantum states of the detected atoms as the strength of the magnetic field was adjusted. 

Recently, in a similar vein, coherent momentum splitting of laser cooled rubidium atoms, prepared in chip-based magnetic traps, was demonstrated using pulsed inhomogeneous magnetic fields  and a sequence of radio-frequency (RF) pulses to prepare, and coherently manipulate, ground-state Zeeman sublevels~\cite{machluf13a}. In these experiments a $\pi/2$ pulse of RF radiation was first applied to prepare a coherent superposition of sublevels with magnetic dipole moments of equal magnitude but opposite orientation, in a homogeneous background magnetic field. The atoms were then subjected to a pulsed inhomogeneous magnetic field. The forces exerted by this field gradient on each component of the internal-state superposition resulted in the generation of a superposition of momentum states. Spatial interferences between the resulting pairs of matter waves were then observed by state-selective imaging of the atoms by laser induced fluorescence. 

The Rydberg-atom interferometer presented here represents an electric analogue of the longitudinal Stern-Gerlach interferometer. It has been implemented using a sequence of microwave pulses to coherently prepare and manipulate selected pairs of Rydberg states. These are interspersed with a pair of electric field gradient pulses to generate spatially separated momentum components. In this scheme a first $\pi/2$ microwave pulse is used to create a superposition of two circular Rydberg states in helium before inhomogeneous electric fields, in conjunction with a $\pi$ pulse of microwave radiation, are applied to generate a spatial separation between the two momentum components associated with this superposition. A second $\mathrm{\pi/2}$ microwave pulse then projects the internal states back onto the circular basis states before state-selective detection of the atoms is carried out by pulsed electric field ionisation.

The remainder of this article is structured as follows: In Sec.~\ref{sec:int_scheme} the Rydberg-atom interferometry scheme used in the experiments is presented. The apparatus and methodologies used to implement this scheme are then described in Sec.~\ref{sec:experiment}. Following this, the results of a set of microwave spectroscopic studies conducted to characterise the Rydberg states and the electric and magnetic fields used in the experiments are outlined in Sec.~\ref{sec:mw_spectroscopy} before the results of the Rydberg-atom interferometry experiments are described in Sec.~\ref{sec:results}. Finally, in Sec.~\ref{sec:conclusion} conclusions are drawn and prospects for future studies briefly discussed.

\section{Interferometry Scheme}\label{sec:int_scheme}

The interferometry scheme implemented in the experiments reported here relies on the creation of a superposition of two Rydberg states with different (induced) electric dipole moments, $\vec{\mu}_{\mathrm{elec,}i}$. In this case we use the $n = 55$ and $n = 56$ circular Rydberg states denoted $|55\mathrm{c}\rangle$ and $|56\mathrm{c}\rangle$ . Beginning with an atom in the $|55\mathrm{c},\vec{p}_0\rangle$ state, where $\vec{p}_0=m\vec{v}_0$ is the initial momentum of the atom with velocity $\vec{v}_0$ and mass $m$, a pulse of microwave radiation resonant with the $|55\mathrm{c}\rangle \rightarrow |56\mathrm{c}\rangle$ transition is applied for a duration such as to cause a coherent $\mathrm{\pi}/2$ rotation of the Rydberg-state population on the Bloch sphere. This operation results in the generation of an equal coherent superposition of the $|55\mathrm{c},\vec{p}_0\rangle$ and $|56\mathrm{c},\vec{p}_0\rangle$ states.

\begin{figure}
	\begin{center}
		\includegraphics[width = 0.65\textwidth, angle = 0, clip=]{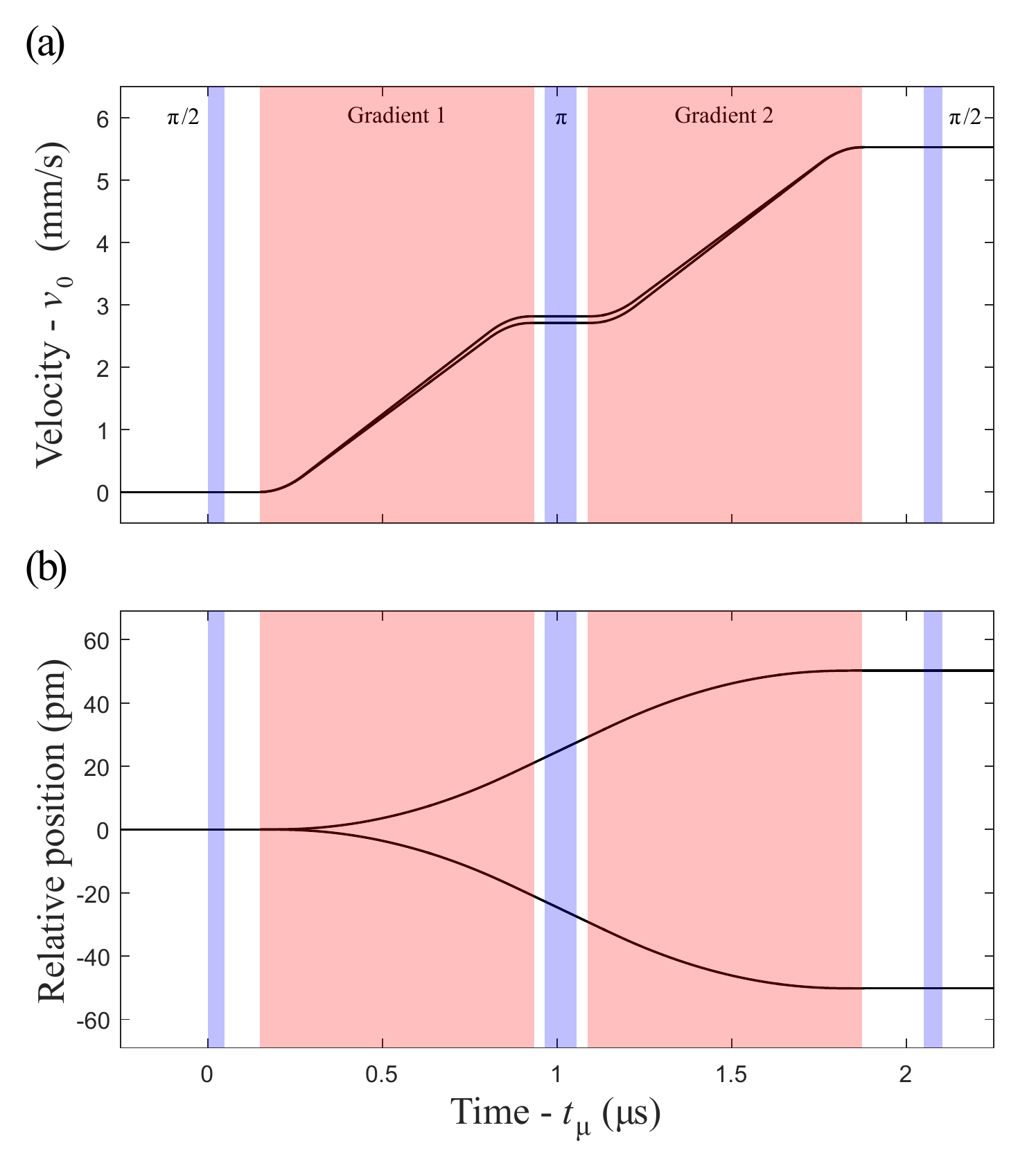}
		\caption{Calculated classical phase-space trajectories of helium atoms in the Rydberg-atom interferometer. (a) The velocity, and (b) the relative position in the $y$ dimension of the trajectories demonstrating equal momenta and a fixed spatial separation at the end of the interferometry procedure. For these calculations $v_0=2000$~m/s, $V_{\mathrm{grad}}$ = 1.3~V, $T_{\mathrm{grad}}=525$ns and $\tau_{\mathrm{grad}}=130$~ns (see text for details).}
		\label{z_v_sim}
	\end{center}
\end{figure}

Atoms in this internal-state superposition are then subjected to a pulsed electric field gradient, $\nabla \vec{F}$. In this field gradient each component of the superposition experiences a force proportional to its electric dipole moment and hence accelerates. This state-dependent acceleration is given by $\vec{a}_i = \vec{\mu}_{\mathrm{elec,}i} \cdot \nabla \vec{F}/m$. Since the two internal states have different (induced) electric dipole moments they accelerate by a different amount and when the gradient pulse is switched off two matter-wave components with different momenta are generated as can be seen, for example, from the classical phase-space trajectories in Fig.~\ref{z_v_sim}(a). Over time, this momentum difference causes a relative displacement between the two matter waves [Fig.~\ref{z_v_sim}(b)]. At the end of this stage of the interferometry procedure the quantum state of the system may therefore be described as an equal coherent superposition of the $|55\mathrm{c},\vec{p}_0 + \mathrm{d}\vec{p}_{55\mathrm{c}}\rangle$ and $|56\mathrm{c},\vec{p}_0 + \mathrm{d}\vec{p}_{56\mathrm{c}}\rangle$ states, where $\mathrm{d}\vec{p}_{i}$ is the state-dependent change in momentum.

Next, a second microwave pulse is applied to induce a $\mathrm{\pi}$ rotation on the Bloch sphere. This inverts the atomic population, i.e., $|55\mathrm{c},\vec{p}_0 + \mathrm{d}\vec{p}_{55\mathrm{c}}\rangle \rightarrow |56\mathrm{c},\vec{p}_0 + \mathrm{d}\vec{p}_{55\mathrm{c}}\rangle$ and $|56\mathrm{c},\vec{p}_0 + \mathrm{d}\vec{p}_{56\mathrm{c}}\rangle \rightarrow |55\mathrm{c},\vec{p}_0 + \mathrm{d}\vec{p}_{56\mathrm{c}}\rangle$. The atoms are then subjected to a second pulsed electric field gradient identical to the first. This combination of $\mathrm{\pi}$ and gradient pulses has the effect of transforming the momenta of the two components of the superposition state to be equal since the component that was accelerated more (less) by the first electric field gradient is accelerated less (more) by the second. The resulting quantum state of the system then becomes an equal superposition of the $|56\mathrm{c},\vec{p}_0 + \mathrm{d}\vec{p}_{55\mathrm{c}} + \mathrm{d}\vec{p}_{56\mathrm{c}}\rangle =|56\mathrm{c},\vec{p}_1\rangle$ and $|55\mathrm{c},\vec{p}_0 + \mathrm{d}\vec{p}_{56\mathrm{c}} + \mathrm{d}\vec{p}_{55\mathrm{c}}\rangle=|55\mathrm{c},\vec{p}_1\rangle$ states, where $\vec{p}_1 = \vec{p}_0 + \mathrm{d}\vec{p}_{55\mathrm{c}} + \mathrm{d}\vec{p}_{56\mathrm{c}}$. At this stage in the sequence of pulses the displacement of the centres of mass of the two matter wave components is fixed since they have equal momenta (see Fig.~\ref{z_v_sim}). They also have equal de Broglie wavelengths which allows for the observation of high contrast interference.

To complete the interferometry scheme a final $\mathrm{\pi}/2$ microwave pulse is applied to project the final internal-state superposition onto the $|55\mathrm{c}\rangle$ and $|56\mathrm{c}\rangle$ basis states. Since the internal Rydberg states of the atoms are entangled with the external motional states at the end of this interferometry procedure, selective detection of the populations of these two circular Rydberg states allows the effects of mater-wave interference to be observed. The final spatial separation of the two matter-wave components is governed by the magnitude and duration of the electric field gradient pulses, and the free evolution time between them. Rydberg-atom interference fringes can therefore be seen by adjusting these parameters.

\section{Experiment}\label{sec:experiment}

\begin{figure}
	\begin{center}
		\includegraphics[width = 0.85\textwidth, angle = 0, clip=]{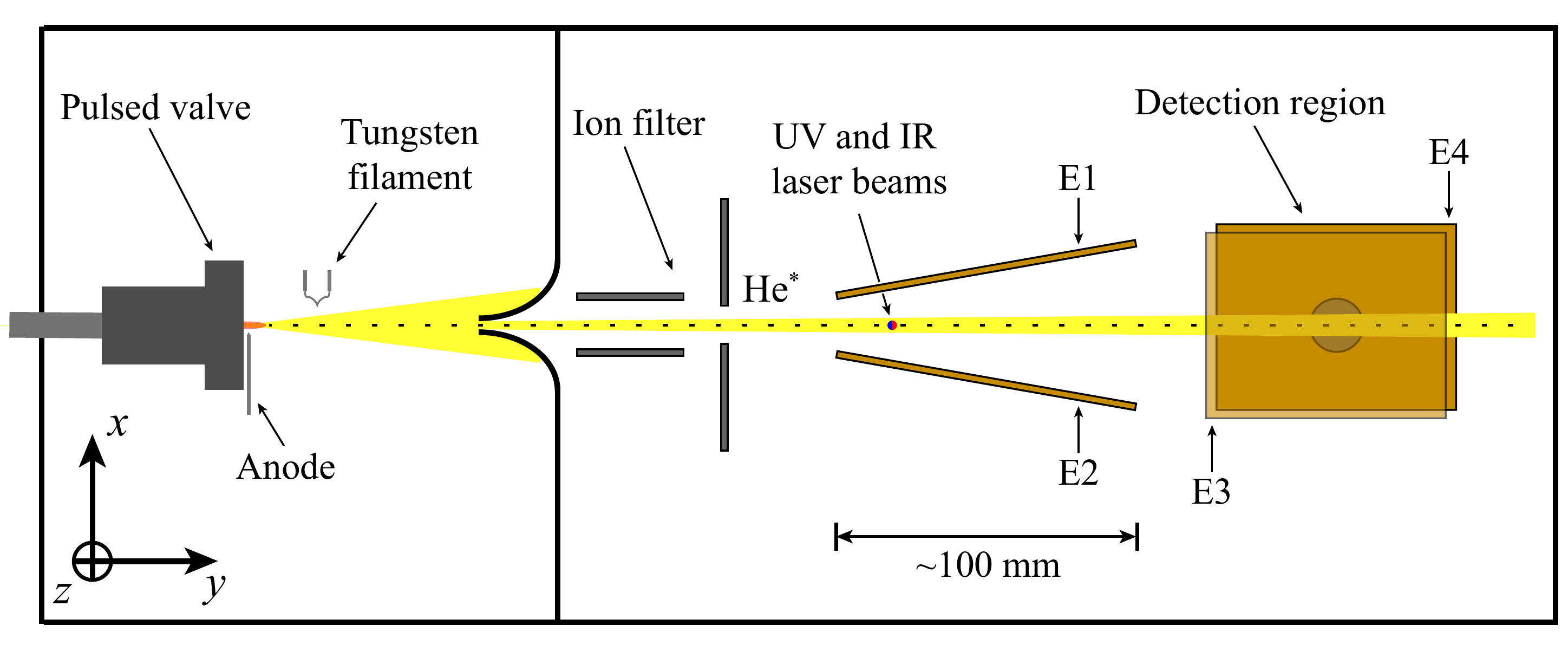}
		\caption{Schematic diagram of the experimental apparatus. Rydberg atom photoexcitation and interferometry were performed within the wedge structure formed by electrodes E1 and E2. State-selective detection by pulsed electric field ionisation occurred between electrodes E3 (partially transparent in the figure) and E4. The MCP detector, to which the ionised electrons were accelerated, was located behind E4.}
		\label{experiment}
	\end{center}
\end{figure}

\begin{figure}
	\begin{center}
		\includegraphics[width = 0.48\textwidth, angle = 0, clip=]{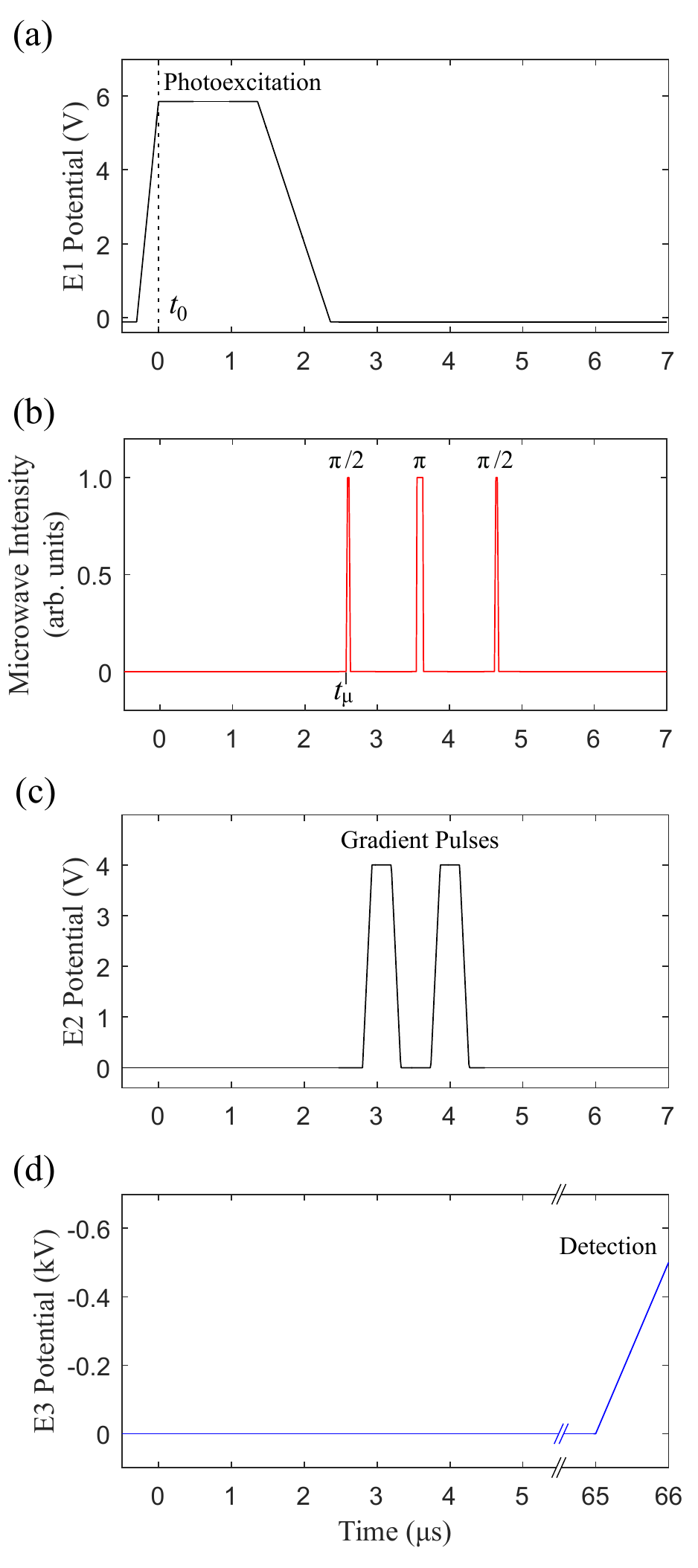}
		\caption{Sequence of pulsed electric potentials and microwave fields employed for Rydberg-atom interferometry. (a) Pulsed potential applied to electrode E1 for Rydberg-state photoexcitation and circular state preparation. (b) Set of three pulses of microwave radiation resonant with the $|55\mathrm{c}\rangle\rightarrow|56\mathrm{c}\rangle$ transition employed for coherent internal state preparation and manipulation. (c) Pair of pulsed potentials applied to electrode E2 to generate the electric field gradients required for interferometry. (d) Slowly-rising pulsed potential applied to electrode E3 for detection by state-selective electric field ionisation. }
		\label{pulses}
	\end{center}
\end{figure}

Fig.~\ref{experiment} shows a schematic diagram of the experimental setup. A pulsed supersonic beam of metastable helium atoms was generated in an electric discharge at the exit of a pulsed valve~\cite{halfmann00a}. The mean velocity of the atoms in this beam was $\vec{v}_0$= (0,2000,0)~m/s. After passing through a 2-mm-diameter skimmer, ions from the discharge were filtered and the atoms entered into a wedge shaped pair of 105 $\times$ 70~mm copper electrodes labelled E1 and E2 in Fig.~\ref{experiment}. These electrodes were separated by 11.5~mm (35.8~mm) in the $x$ dimension at the end nearest to the ion filter (detection region). Hence, for a potential difference of $V_{\mathrm{grad}}$ between them, the electric field gradient on the axis along which the atomic beam propagated was $\nabla \vec{F}=V_{\mathrm{grad}}$ (0,0.056,0) V/cm$^2$. At the position between these electrodes where their separation in the $x$ dimension was 19~mm the atoms were prepared in the $| n = 55, \ell = 54, m_{\ell} =+54 \rangle \equiv |55\mathrm{c}\rangle$ circular Rydberg state using the crossed fields method~\cite{delande88a,zhelyazkova16a}. To implement this, laser photoexcitation was carried out in the presence of perpendicular pulsed electric and static magnetic fields of (3.15,0,0)~V/cm and (0,0,15.915)~G, respectively. In this process the atoms were excited from the metastable 1s2s\,$^3\mathrm{S}_1$ level to the intermediate 1s3p\,$^3\mathrm{P}_2$ level and then the outer $n = 55$ Rydberg-Stark state using focused, co-propagating cw lasers in the ultraviolet ($\lambda_{\mathrm{uv}}= 388.975~\mathrm{nm} \equiv 25708.6$~cm$^{-1}$) and infrared ($\lambda_{\mathrm{ir}} = 786.752~\mathrm{nm} \equiv 12710.5$~cm$^{-1}$) regions of the electromagnetic spectrum~\cite{hogan18a}. After photoexcitation the electric field was switched off adiabatically to transfer the atomic population into the $|55\mathrm{c}\rangle$ state as indicated in Fig.~\ref{pulses}(a).

When circular state preparation was completed, the sequence of microwave and electric field gradient pulses used to implement the Rydberg-atom interferometry scheme described in Sec.~\ref{sec:int_scheme} was applied. This sequence was initiated at time $t_\mathrm{\mu} = t_0 + 2.575~\mu$s [see Fig.~\ref{pulses}(b)]. At the end of this sequence, which lasted 2.1~$\mu$s, the atoms travelled for $\sim$60~$\mu$s [see Fig.~\ref{pulses}(d)] to the detection region of the apparatus, $\sim130$~mm downstream from the position of laser photoexcitation, where they were ionised by applying a slowly-rising potential of -530~V to E3 generating fields of up to $\sim$115~V/cm directed parallel to the background magnetic field. The resulting electrons were accelerated through an aperture in E4 to a microchannel plate (MCP) detector. Since the $|56\mathrm{c}\rangle$ state ionises in a lower field than the $|55\mathrm{c}\rangle$ state, electrons from atoms in this state appeared earlier in the time-of-flight distributions measured at the MCP. By selecting appropriate windows in the time-of-flight distributions, the populations of these two circular states of interest could be determined.

Circular Rydberg states were used in the experiments because the large electric dipole moments for transitions between them allowed fast state preparation and manipulation with pulsed microwave fields. These states also have a low sensitivity to weak stray electric fields and electric field noise resulting in minimal decoherence under these conditions. Finally, since blackbody transitions from these states are significantly restricted by the electric dipole selection rules, changes in their population between the end of the interferometry pulse sequence and the time when they were detected by pulsed electric field ionisation was minimal.

\section{Microwave Spectroscopy}\label{sec:mw_spectroscopy}

\begin{figure}
	\begin{center}
		\includegraphics[width = 0.48\textwidth, angle = 0, clip=]{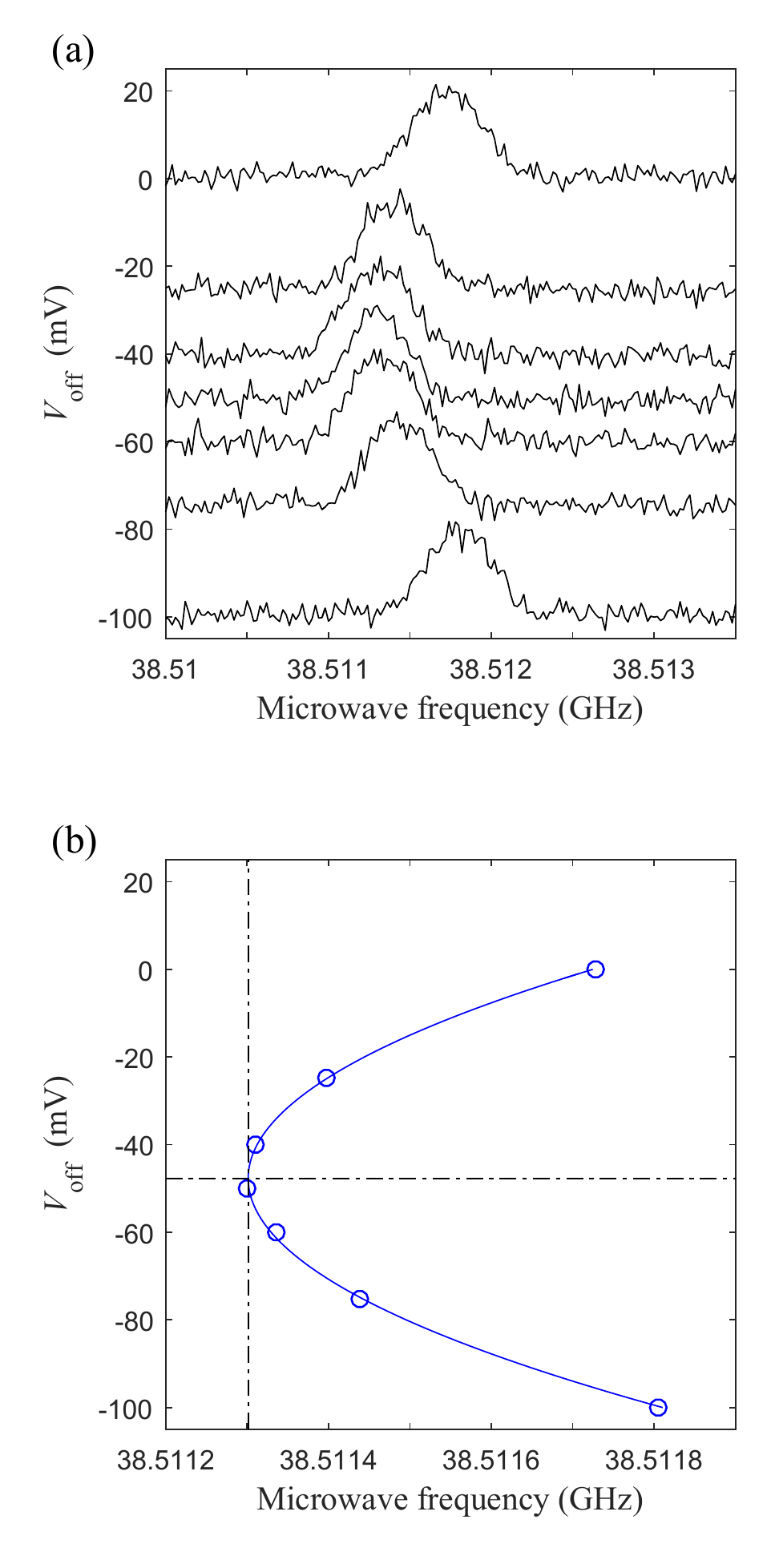}
		\caption{(a) Microwave spectra of the $|55\mathrm{c}\rangle \rightarrow |56\mathrm{c}\rangle$ transition showing the Stark shifts for offset potentials, $V_{\mathrm{off}}$, applied to E2. In this panel the vertical offset of each spectrum indicates the value of $V_{\mathrm{off}}$ for which it was measured. (b) The quadratic function (continuous curve) fit to the measured transition frequencies obtained from the data in panel (a). The zero-electric-field transition frequency of $\nu_{\mathrm{\mu}}$ = 38.511313~GHz for an offset potential of $V_{\mathrm{off}}$ = -48~mV is indicated by the dashed lines in panel (b).}
		\label{stark_shifts}
	\end{center}
\end{figure}

High-resolution microwave spectroscopic studies of the $|55\mathrm{c}\rangle \rightarrow |56\mathrm{c}\rangle$ transition were carried out to calibrate and optimise the parameters required to implement the Rydberg-atom interferometry scheme described in Sec.~\ref{sec:int_scheme}. For the circular Rydberg states used in the experiments the Stark energy shifts in an electric field $\vec{F}=(F_x,0,0)$ acting in the direction perpendicular to the background magnetic field $\vec{B}=(0,0,B_z)$ can be expressed as~\cite{delande88a,pauli26a}

\begin{equation}\label{eq:StarkZeeman}
E_{n\mathrm{c}}= m_\ell \sqrt{\bigg(\mu_\mathrm{B} B_z\bigg)^2+\bigg(\frac{3}{2} n e a F_x\bigg)^2},
\end{equation}
where $m_\ell$ is defined with respect to the magnetic field quantisation axis,  and $\mu_\mathrm{B}$, $e$ and $a$ are Bohr magneton, electron charge and Bohr radius corrected for the reduced mass, respectively. In weak electric fields, i.e., when the electric field term in Eq.~\ref{eq:StarkZeeman} is smaller than the magnetic field term, this expression results in quadratic Stark shifts of the circular Rydberg states. As the field strength is increased, the electric field term begins to dominate that associated with the magnetic field and the Stark energy shifts become increasingly linear. 

To compensate the $(32,0,0)$~mV/cm motional Stark effect~\cite{elliott95a} and weak stray electric fields in the apparatus, microwave spectra of the $|55\mathrm{c}\rangle \rightarrow |56\mathrm{c}\rangle$ transition were recorded for a range of offset potentials, $V_{\mathrm{off}}$, applied to E1. In recording these spectra microwave pulses of $2~\mu$s duration were applied at time $t_{\mu}$ [see Fig.~\ref{pulses}(b)]. The resulting data are presented in Fig.~\ref{stark_shifts}(a) where the vertical offset of each spectrum corresponds to the value of $V_{\mathrm{off}}$. From the resonance frequencies in each of these spectra, determined by fitting Gaussian functions to each dataset, and the corresponding value of $V_{\mathrm{off}}$, the optimal compensation potential was determined by fitting a function that depended quadratically on $V_{\mathrm{off}}$. The results of this procedure are displayed in Fig.~\ref{stark_shifts}(b) and yielded a minimum, i.e., a zero-electric-field, transition frequency of $\nu_{\mathrm{\mu}}=38.511\,313$~GHz for $V_{\mathrm{off}}=-48$~mV. From this transition frequency, the strength of the magnetic field at the position of the atoms in the apparatus was determined, using the Rydberg formula and the expression in Eq.~\ref{eq:StarkZeeman}, to be $B_{\mathrm{z}}$ = 15.915~G.

\begin{figure}
	\begin{center}
		\includegraphics[width = 0.65\textwidth, angle = 0, clip=]{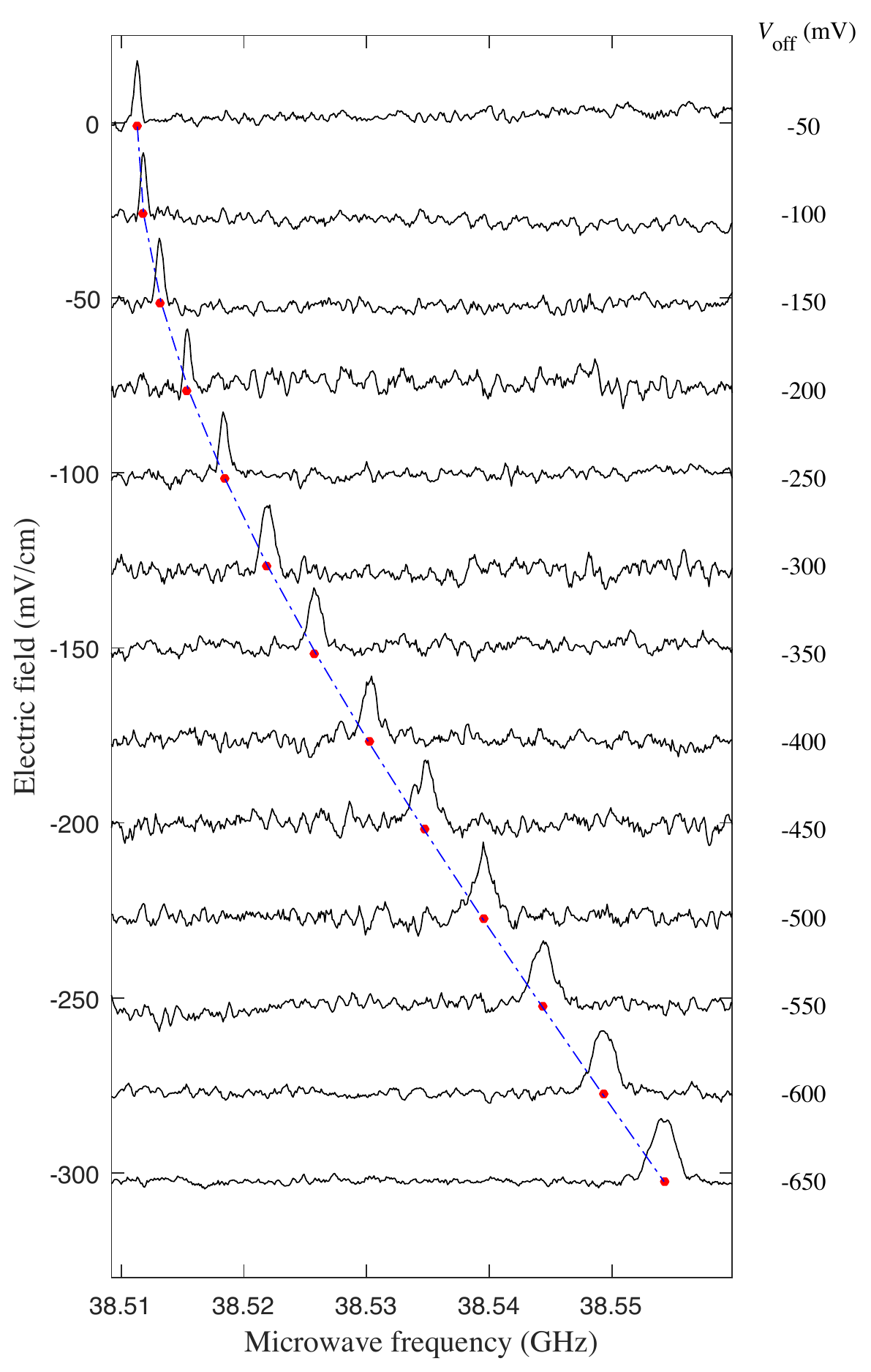}
		\caption{Microwave spectra of the $|55\mathrm{c}\rangle \rightarrow |56\mathrm{c}\rangle$ transition in a magnetic field $B_{\mathrm{z}}$ = 15.915~G and perpendicular electric fields as indicated on the left vertical axis. The corresponding values of the offset potential, $V_{\mathrm{off}}$, applied to E1 are indicated on the right. The dashed curve represents the transition frequencies calculated using Eq.~\ref{eq:StarkZeeman}. The red points correspond to the experimentally determined transition frequencies.}
		\label{stark_shifts_2}
	\end{center}
\end{figure}

The accuracy of the hydrogenic expression in Eq.~\ref{eq:StarkZeeman} for describing the Stark shifts of  $|55\mathrm{c}\rangle$ and $|56\mathrm{c}\rangle$ states of helium in weak perpendicular fields was tested by comparing experimentally recorded spectra with the calculated transition frequencies over a wider range of electric fields. The results of these measurements are displayed in Fig.~\ref{stark_shifts_2}. In recording these spectra the microwave intensity was set to ensure that $<10\%$ population transfer from the $|55\mathrm{c}\rangle$ state to the $|56\mathrm{c}\rangle$ state occurred. The spectral width of the transition in the measurement made in the electric field closest to zero (see left axis), for which $V_{\mathrm{off}}$ = -48~mV, is 590~kHz and corresponds approximately to the Fourier Transform limit of the 2~$\mu$s duration microwave pulse. For the larger electric fields in which these spectra were recorded the spectral features are broader. This is because in these fields the atoms are more strongly polarised and therefore more sensitive to the inhomogeneity in the electric field and electric field noise~\cite{zhelyazkova15a,zhelyazkova15b}. The dashed curve and red points overlaid on the experimental spectra in this figure represent the calculated and measured transition frequencies, respectively, and exhibit excellent quantitative agreement. From Eq.~\ref{eq:StarkZeeman} the maximal induced electric dipole moments of the $|55\mathrm{c}\rangle$ and $|56\mathrm{c}\rangle$ states was determined to be 11\,320~D and 11\,740~D, respectively. The difference in these dipole moments converges to its maximal value of 420~D in fields above $\sim0.6$~V/cm. 

\begin{figure}
	\begin{center}
		\includegraphics[width = 0.4\textwidth, angle = 0, clip=]{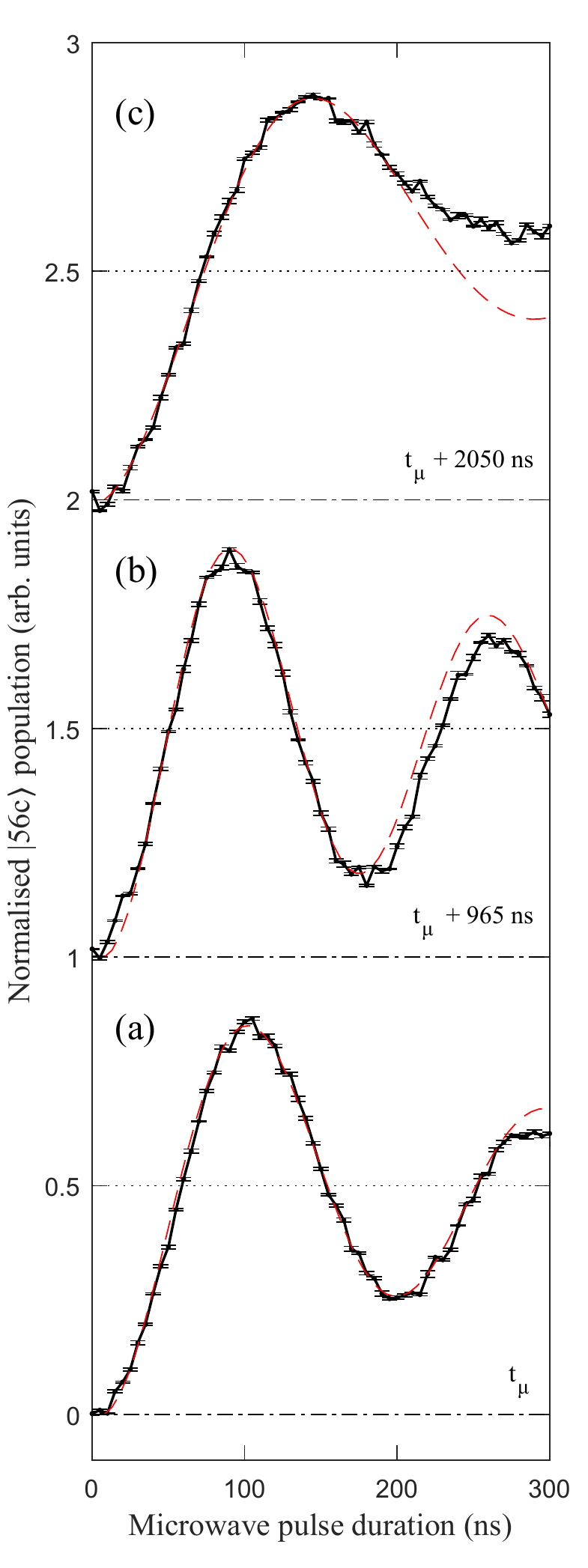}
		\caption{Rabi oscillations in the population of the $|56\mathrm{c}\rangle$ state upon resonant driving of the $|55\mathrm{c}\rangle \rightarrow |56\mathrm{c}\rangle$ transition for a fixed output power from the microwave source. Measurements were performed at the time delays of (a) $t_{\mathrm\mu}$, (b) $t_{\mathrm\mu}+965$~ns and (c) $t_{\mathrm\mu}+2050$~ns at which each of the three microwave pulses required to implement the Rydberg-atom interferometry scheme detailed in Sec.~\ref{sec:int_scheme} was applied (see Fig~\ref{pulses}).}
		\label{rabi}
	\end{center}
\end{figure}

To determine the appropriate length of time to apply microwave pulses to realise the three coherent rotations on the Bloch sphere required to implement the interferometry scheme described in Sec.~\ref{sec:int_scheme}, i.e., a $\mathrm{\pi/2}$ pulse to create an internal-state superposition, a $\mathrm{\pi}$ pulse to invert the states, and a final $\mathrm{\pi/2}$ pulse to project back onto the basis states, the $|55\mathrm{c}\rangle$ state was prepared as above and pulses of microwave radiation at the resonant frequency $\nu_{\mathrm{\mu}}$ = 38.511\,313~GHz were applied at a range of times, and with a range of durations. The results of these measurements can be seen in Fig.~\ref{rabi} as Rabi oscillations in the population of the $|56\mathrm{c}\rangle$ state. The time delays of $t_{\mathrm{\mu}} $, $t_{\mathrm{\mu}} + 965$~ns and $t_{\mathrm{\mu}} + 2050$~ns indicated in this figure correspond to the timings of the three microwave pulses in the interferometry measurements. From these results it was determined that the pulse durations required were $\tau_{\mathrm{\pi/2}}$ = 48~ns for the first $\mathrm{\pi/2}$ pulse, $\tau_{\mathrm{\pi}}$ = 88~ns for the $\mathrm{\pi}$ pulse, and $\tau_{\mathrm{\pi/2}}$ = 52~ns for the second $\mathrm{\pi/2}$ pulse. The inconsistency in the durations of these pulses, particularly the two $\mathrm{\pi/2}$ pulses, is a consequence of the mode structure of the microwave field within the wedge configuration between E1 and E2. The periodicity and decoherence rates of the Rabi oscillations vary for the different time delays because of the inhomogeneity of the stray electric fields in the electrode structure. At each time delay the atoms are in a different location ($\sim$4~mm difference between $t_{\mathrm{\mu}}$ and $t_{\mathrm{\mu}}+2050$~ns) and therefore experience  slightly different electric fields that shift the atomic transition frequency away from resonance with the microwave field.

\begin{figure}
	\begin{center}
		\includegraphics[width = 0.5\textwidth, angle = 0, clip=]{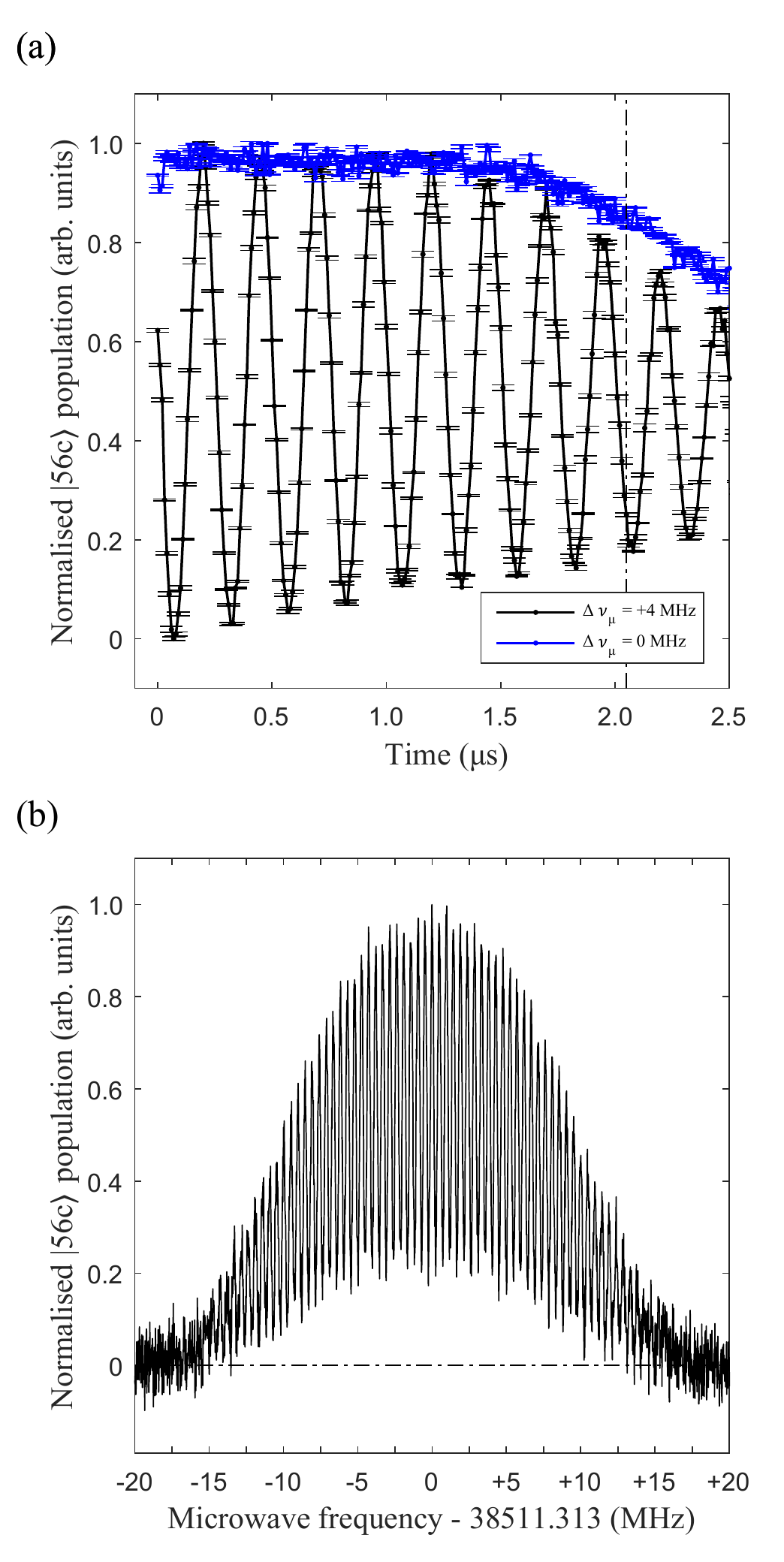}
		\caption{(a) Ramsey interferogram, and (b) Ramsey spectrum of the $|55\mathrm{c}\rangle \rightarrow |56\mathrm{c}\rangle$ transition. In (a) measurements performed for detunings of the microwave field from resonance of $\Delta\nu_{\mathrm{\mu}}$ = 0~MHz and $\Delta\nu_{\mathrm{\mu}}$ = +4~MHz are indicated by the blue and black data sets, respectively. The separation of 2050~ns between the application of the two $\mathrm{\pi/2}$ microwave pulses used ultimately for the Rydberg-atom interferometry experiments is indicated by the dashed vertical line. The spectrum in (b) was recorded for microwave pulses of the same duration as those used in (a) but for a fixed free-evolution time of 2002~ns.}
		\label{ramsey_both}
	\end{center}
\end{figure}

The coherence of the circular state superpositions prepared in the experiments was characterised by Ramsey interferometry and spectroscopy. For these measurements, the results of which are displayed in Fig.~\ref{ramsey_both}, only the 48- and 52-ns-duration $\mathrm{\pi/2}$ pulses were applied. In the Ramsey interferometry measurement seen in Fig.~\ref{ramsey_both}(a) the time interval between these pulses was adjusted as the population of the $|56\mathrm{c}\rangle$ state was monitored. A first measurement performed on resonance, i.e., for $\Delta\nu_{\mathrm{\mu}}$ = 0~MHz, yielded an approximately constant population of this state for all time intervals between the $\mathrm{\pi/2}$ pulses up to $\sim 1.5~\mu$s. After this the population reduced because of effects of decoherence and dephasing. In this set of data no oscillations in the $|56\mathrm{c}\rangle$ state population are observed because of the continuous evolution of the phase of the gated microwave field in the time between the two pulses. On the other hand, for a detuning of $\Delta\nu_{\mathrm{\mu}}$ = +4~MHz periodic oscillations in the population of the $|56\mathrm{c}\rangle$ state are seen. These occur, as expected, at the frequency of 4~MHz associated with the detuning, and reduce in amplitude in a similar way to the data recorded on resonance. From these data the decoherence times of these states were determined to be $\sim5~\mu$s, limited by the motion of the atoms. Because of the onset of more significant decoherence for times beyond $2.05~\mu$s [dashed line in Fig~\ref{ramsey_both}(a)] this time interval was chosen as the maximal length for the sequence of pulses used in the interferometry experiments.

A Ramsey spectrum of the $|55\mathrm{c}\rangle \rightarrow |56\mathrm{c}\rangle$ transition recorded in the frequency domain with the same pair of $\mathrm{\pi/2}$ microwave pulses separated by a free-evolution time of 2002~ns, corresponding to the application of the second $\mathrm{\pi/2}$ at time $t_{\mathrm{\mu}} + 2050$~ns, is displayed in Fig.~\ref{ramsey_both}(b). This spectrum further demonstrates the coherence of the atom--microwave-field coupling over the time scales of interest in the experiments. It also indicates an $\sim$80\% fidelity of this Ramsey sequence close to resonance and the accurate identification of the zero-electric-field resonance frequency for the transition.

\begin{figure*}
	\begin{center}
		\includegraphics[width = 1\textwidth, angle = 0, clip=]{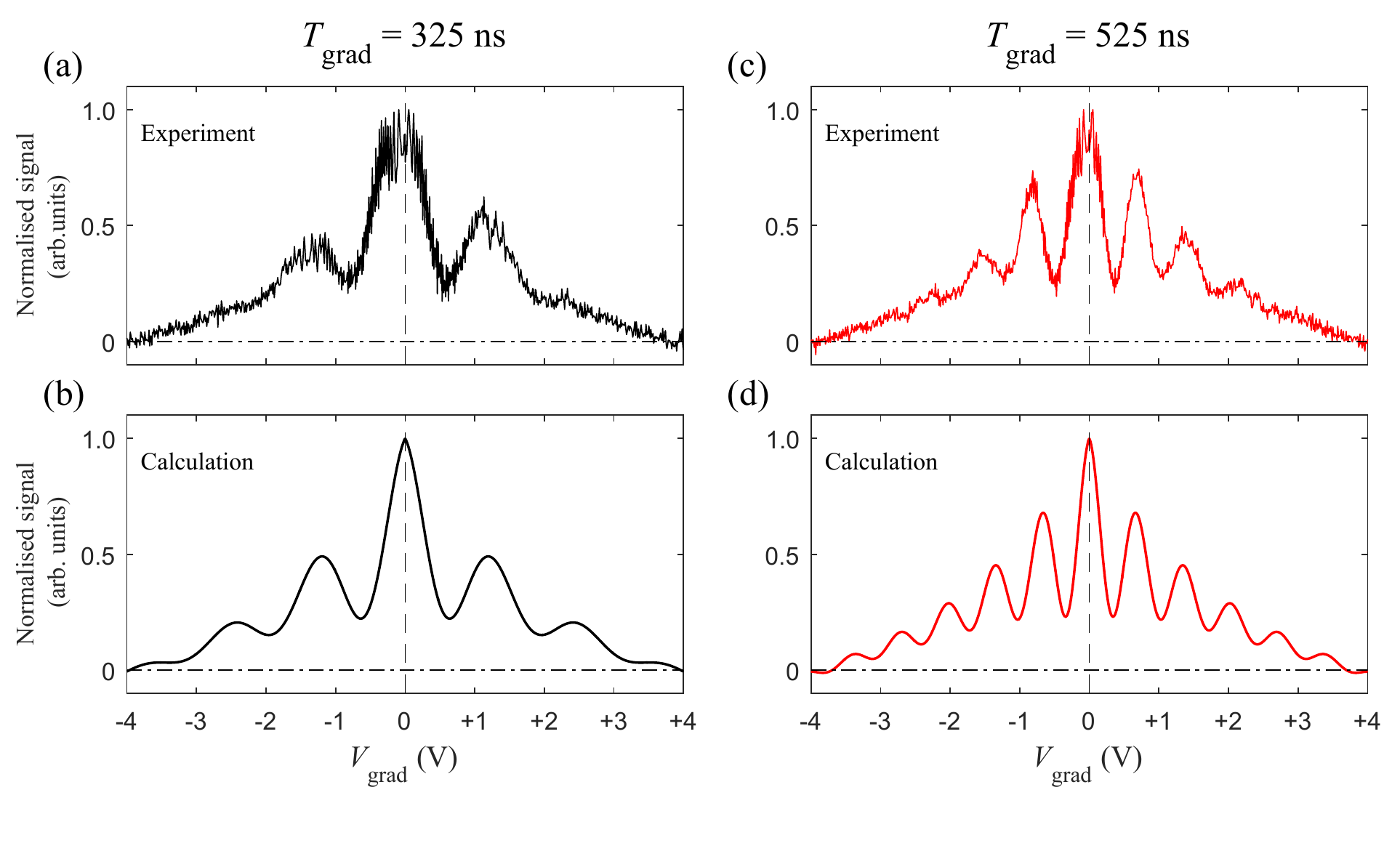}
		\caption{(a) and (c) measured, and (b) and (d) calculated Rydberg-atom interference patterns obtained for pairs of electric field gradient pulses with durations $T_{\mathrm{grad}}=325$ and 525~ns, respectively. The amplitude of the electric potential, $V_{\mathrm{grad}}$, applied to electrode E2 to generate the field gradients is indicated on the horizontal axis. In all cases, $\tau_{\mathrm{grad}}=130$~ns.}
		\label{fringes}
	\end{center}
\end{figure*}

\section{Results}\label{sec:results}

Having characterised each of the elements of the sequence of microwave and electric field gradient pulses required for Rydberg-atom interferometry, a series of experiments were performed to study the operation of the interferometer. In these experiments, the relative populations of the Rydberg states at the end of the interferometry procedure reflected the spatial overlap of the de~Broglie waves associated with the two components of the superposition of momentum states at the time the second $\mathrm{\pi/2}$ microwave pulse was applied (see Fig.~\ref{pulses}). This spatial separation was dependent on both the amplitude and duration of the electric field gradient pulses. Fig.~\ref{fringes}(a) and~(c) show interference patterns measured by adjusting the amplitudes of the pulsed electric potentials applied to generate the gradient pulses for two different gradient pulse durations, $T_{\mathrm{grad}}=325$ and 525~ns. These durations are the length of time for which the potentials applied to generate the gradients were at their maximum value, $V_{\mathrm{grad}}$. In these cases each gradient pulse had rise and fall times of $\tau_{\mathrm{grad}}$ = 130~ns. For each of these measurements the ratio of the $|55\mathrm{c}\rangle$-electron signal to that from the $|56\mathrm{c}\rangle$ state is displayed and normalised to one for the maximum measured signal in each data set. The signal maximum for values of $|V_{\mathrm{grad}}|$ close to zero indicates that the Rydberg-state population at the end of the interferometry sequence resided predominantly in the $|55\mathrm{c}\rangle$ state. For the short gradient pulse of 325~ns [Fig.~\ref{fringes}(a)], matter-wave interference begins to become observable for values of $|V_{\mathrm{grad}}|\gtrsim0.2$. The first interference minimum, when the centres-of-mass of the two de~Broglie wavepackets associated with the two Rydberg-atom momentum components are displaced by $\lambda_{\mathrm{dB}}/2$ at the time when the last $\pi/2$ microwave pulse was applied, is then observed for $|V_{\mathrm{grad}}|\simeq0.8$. When the amplitude of the pulsed gradients was further increased the larger forces exerted on the atoms further displace the momentum components from each other with a displacement of the corresponding de~Broglie waves of $\lambda_{\mathrm{dB}}$ when $|V_{\mathrm{grad}}|\simeq1.2$. The non-linear dependence of the values of $|V_{\mathrm{grad}}|$ at which these interference features occur is a result of the quadratic Stark shifts of the circular Rydberg states in weak electric fields (see Eq.~\ref{eq:StarkZeeman} and Fig.~\ref{stark_shifts_2}). For higher values of $|V_{\mathrm{grad}}|$ the contrast of the measured interference pattern reduces. A similar general behaviour is seen in Fig.~\ref{fringes}(c) recorded for $T_{\mathrm{grad}}=525$~ns. For this longer pulse duration, since the forces applied to generate the coherent superposition of momentum components are greater, larger matter-wave separations are achieved for each value of $V_{\mathrm{grad}}$. In this case the first interference minimum is observed for $|V_{\mathrm{grad}}|\simeq0.4$~V while the following interference maximum occurs for $|V_{\mathrm{grad}}|\simeq0.7$~V. A greater number of interference fringes are seen in this data than in the data recorded for $T_{\mathrm{grad}}=325$~ns. However, the dependence of the contrast of the observed fringes on the value of $|V_{\mathrm{grad}}|$ remains similar to that in Fig.~\ref{fringes}(a). This suggests that the loss of contrast is not a consequence of the displacement of the matter waves but is instead correlated with the amplitude of the pulsed inhomogeneous electric fields applied. 

To aid in the interpretation of the experimental data in Fig.~\ref{fringes}, numerical calculations of the matter-wave interference patterns were performed. In these calculations the classical equations of motion of atoms in the $|55\mathrm{c}\,\rangle$ and $|56\mathrm{c}\,\rangle$ states were solved in the magnetic and time-dependent inhomogeneous electric fields used in the experiments. These calculations were initiated after circular state preparation. From this time until the time at which the $\pi$ pulse was applied, the forces exerted on each atom in the initial Rydberg state in which it was prepared were determined. At the time of the $\pi$ pulse, the states for which the atomic trajectories were calculated were inverted. The calculations then continued until the end of the microwave pulse sequence. At this time, the spatial separation between the end points of the two atomic trajectories was extracted (see, e.g., Fig.~\ref{z_v_sim}). The interference between two plane waves with this separation and for which $\lambda_{\mathrm{dB}} = h/p_0=50$~pm, where $p_0 =|\vec{p_0}|$, was then obtained. The resulting Rydberg-atom interference patterns in Fig.~\ref{fringes}(b) and~(d) which were calculated for the values of $T_{\mathrm{grad}}$ in Fig.~\ref{fringes}(a) and~(c), respectively, exhibit interference fringes with the same periodicity as those in the experimental data. However, to achieve the excellent quantitative agreement with the results of the experiments seen in this figure it was necessary to incorporate into the calculations contributions from decoherence and population loss from the circular states. Both of these effects depended exponentially on the value of $|V_{\mathrm{grad}}|$ resulting in interference patterns with intensity
\begin{eqnarray}
I(V_{\mathrm{grad}}) &=& A\,\left[I_0(V_{\mathrm{grad}})-0.5\right]\,\exp\left[-\frac{|V_{\mathrm{grad}}|}{C_{\mathrm{decoh.}}}\right] + \exp\left[-\frac{|V_{\mathrm{grad}}|}{C_{\mathrm{loss}}}\right] + B,
\end{eqnarray}
where $I_0(V_{\mathrm{grad}})$ is the dependence of the normalised intensity of the interference pattern on the value of $|V_{\mathrm{grad}}|$, excluding effects of dephasing, decoherence or circular state depopulation, and $A = 0.8$ and $B=-0.4$ are constants. The decay constant, $C_{\mathrm{decoh.}}$, associated with decoherence of the superposition states, which caused the reduction in the contrast of the interference fringes, was found to be 1.2~V. The decay constant, $C_{\mathrm{loss}}$, associated with the loss in the total detected signal from the circular Rydberg states, which caused the reduction in the normalised signals in Fig.~\ref{fringes}(a) and~(c), was found to be 4.2~V.  The decoherence inferred from the results of these calculations is attributed to a combination of the finite longitudinal coherence length of the momentum distribution in the atomic beam, effects of electric field noise on the Rydberg states that increase when they are more strongly polarised, and imperfections in the relative amplitudes of the pairs of electric field gradient pulses. The elucidation of the mechanism that caused the loss in the total signal detected as the value of $|V_{\mathrm{grad}}|$ was increased required the performance of additional experiments. However, from the results of the calculations in Fig.~\ref{fringes}(b) and (d) it was determined that, e.g., when $|V_{\mathrm{grad}}|=1.3$~V in Fig.~\ref{fringes}(c) and the second interference maximum is observed the difference in the forces exerted on the two matter-wave components in the pulsed field gradients was $1.4\times10^{-24}$~N. This corresponds to a difference in acceleration of $220~\mathrm{m/s}^2$ and a displacement of the centres-of-mass of the matter-waves at the time of the second $\mathrm{\pi/2}$ pulse of 100~pm (see also Fig.~\ref{z_v_sim}).

\begin{figure}
	\begin{center}
		\includegraphics[width = 0.55\textwidth, angle = 0, clip=]{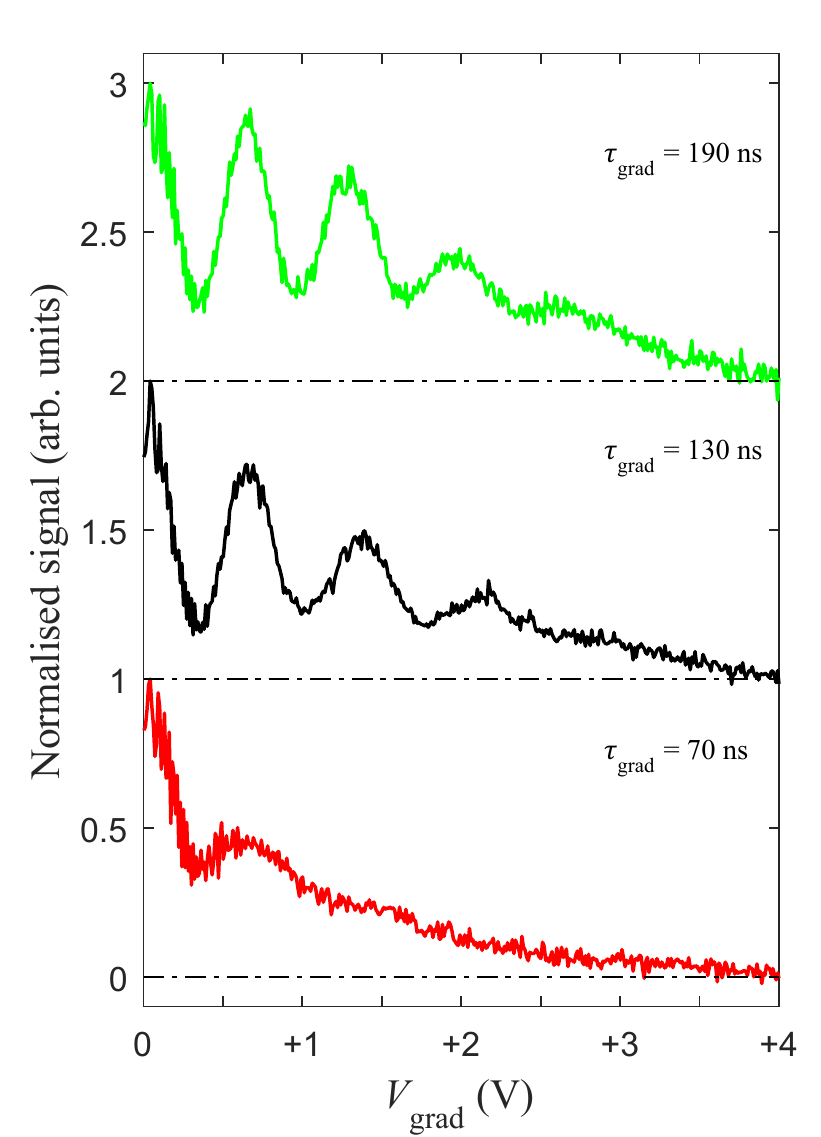}
		\caption{Rydberg-atom interference patterns recorded for $T_{\mathrm{grad}}=525$~ns and values of $\tau_{\mathrm{grad}}$ of 70, 130 and 190~ns, as indicated. For clarity only the parts of each interference pattern for which $V_{\mathrm{grad}}\geq0$ are displayed. The data recorded for $\tau_{\mathrm{grad}}=130$~ns corresponds to Fig.~\ref{fringes}(c). Note the upper datasets in this figure are vertically offset for clarity.}
		\label{edge_times}
	\end{center}
\end{figure}

To identify the origin of the loss of signal at high values of $|V_{\mathrm{grad}}|$ in Fig.~\ref{fringes}, further measurements were performed for $T_{\mathrm{grad}}=525$~ns but for a range of values of $\tau_{\mathrm{grad}}$. Data recorded for $\tau_{\mathrm{grad}}=70$, 130 and 190~ns are displayed in Fig.~\ref{edge_times}. For clarity, only results for $V_{\mathrm{grad}}\geq0$ are included in this figure. From these measurements it can be seen that the greatest contrast in the interference patterns, and the greatest number of interference fringes, occur for the slowest rise and fall times of the electric field gradient pulses, i.e., for $\tau_{\mathrm{grad}}=190$~ns. The data recorded for $\tau_{\mathrm{grad}}=130$~ns, which corresponds to that in Fig.~\ref{fringes}(c), exhibits very similar characteristics with only slightly more significant decoherence and loss in total signal at larger values of $V_{\mathrm{grad}}$. However, for the shortest rise time of $\tau_{\mathrm{grad}}=70$~ns the situation is quite different. Not only is decoherence more apparent, but the total population also decays faster than it does in the measurements with slower rise times. From this more significant loss in total population from the two circular states for short rise and fall times, it is concluded that this contribution to the measurements results from the non-adiabatic evolution of the Rydberg states under these conditions. These non-adiabatic dynamics transfer population out of the $|55\mathrm{c}\rangle$ and $|56\mathrm{c}\rangle$ states and into neighbouring lower $m_{\ell}$ states as the pulsed fields are more rapidly switched from and to zero.

\begin{figure}
	\begin{center}
		\includegraphics[width = 0.45\textwidth, angle = 0, clip=]{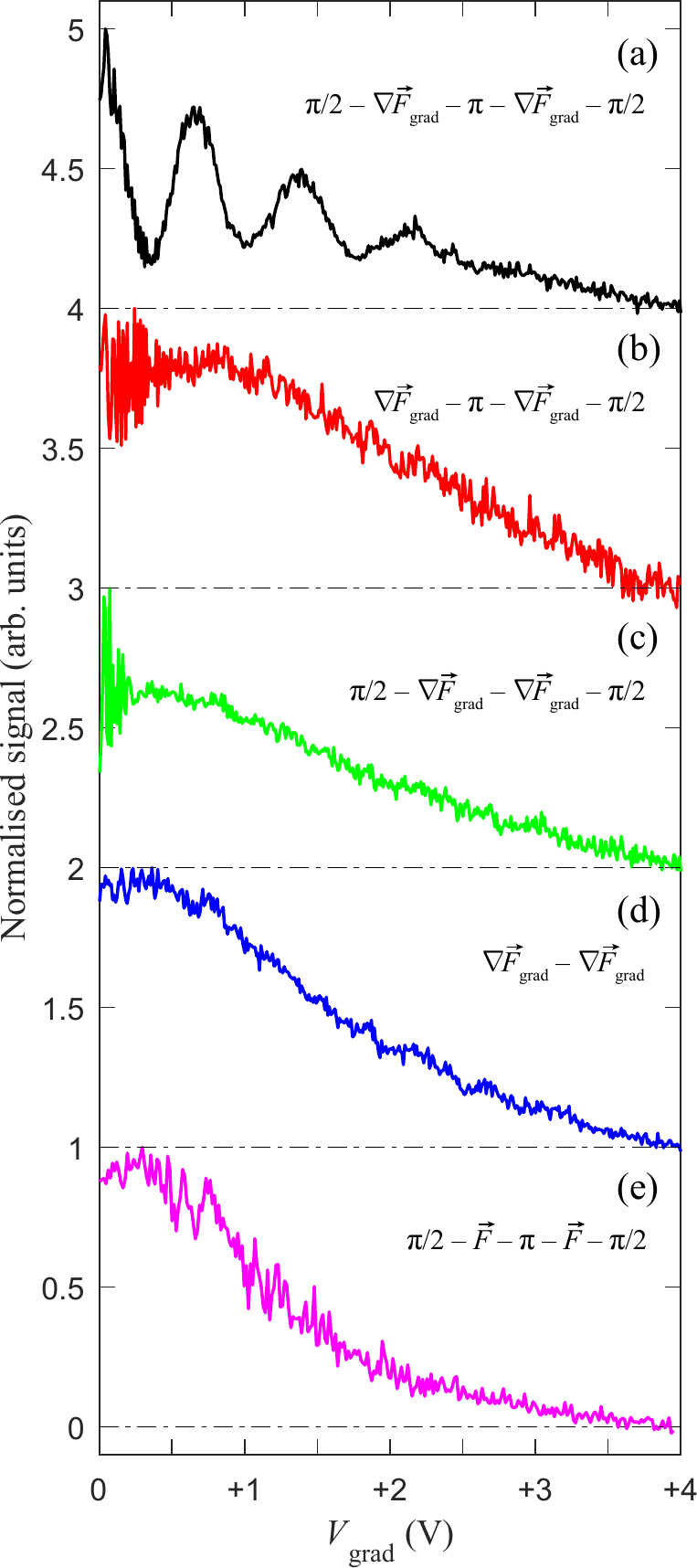}
		\caption{Measurements performed to validate the interpretation of the Rydberg-atom interference patterns. (a) Reference measurement performed with the complete Rydberg-atom interferometry pulse sequence, $T_{\mathrm{grad}}=525$ and $\tau_{\mathrm{grad}}=130$~ns, as in Fig.~\ref{fringes}(c). Interferometry sequences with (a) the first $\pi/2$ microwave pulse, (b) the $\pi$ pulse, and (d) all microwave pulses omitted. (e) Measurement performed with electrodes E1 and E2 oriented parallel to each other. The electric potentials on the horizontal axis in (e) have been scaled by 1.46 so that the corresponding electric fields are comparable to those at the position of the atoms in datasets (a) to (d) (see text for details). }
		\label{test_scans}
	\end{center}
\end{figure}

To confirm the attribution of the oscillations in the data in Fig.~\ref{fringes} to effects of matter-wave interference, and that they are not a consequence of resonances or fluctuations introduced by the pulsed electric fields, several further tests were performed. The results of these are displayed in Fig.~\ref{test_scans}. For all of the measurements in this figure $T_{\mathrm{grad}}=525$~ns and $\tau_{\mathrm{grad}}=130$~ns. The data in Fig.~\ref{test_scans}(a) represents a reference measurement which corresponds to that in Fig.~\ref{fringes}(c). To check that the oscillations in the ratio of the circular state populations upon application of the complete interferometry sequence of microwave and electric field gradient pulses relied on the preparation and coherent manipulation of the superpositions of Rydberg states the data in Fig.~\ref{test_scans}(b), (c) and (d) were recorded with the first $\mathrm{\pi/2}$, the $\mathrm{\pi}$ and all three microwave pulses omitted, respectively. No oscillations are seen with the periodicity of those observed in Fig.~\ref{test_scans}(a) in any of these data sets. The data in Fig.~\ref{test_scans}(d), in which no microwave pulses were applied and the normalised signal gradually reduces as $|V_{\mathrm{grad}}|$ is increased, confirms that the loss in the circular state populations for larger values of $|V_{\mathrm{grad}}|$ is a result of non-adiabatic internal Rydberg-state dynamics and is not caused by decoherence of the superposition of momentum states. 

As a final test, Fig.~\ref{test_scans}(e) shows the result of a measurement carried out with an identical sequence of microwave and electric potential pulses to that in Fig.~\ref{test_scans}(a) but with the configuration of electrodes E1 and E2 adjusted so that they were parallel to each other with a separation in the $x$ dimension of 13~mm. To account for the different electric fields the atoms experience in this configuration for each value of $|V_{\mathrm{grad}}|$, when compared to that in the wedge configuration, and to allow for a more direct comparison with the data in panels (a) to (d), the potentials on the horizontal axis in Fig.~\ref{test_scans}(e) have been scaled by 1.46. In this situation the applied pulsed electric potentials did not give rise to significant electric field gradients at the position of the atoms and therefore no forces were exerted on the atoms by these fields. No oscillations are observed in this data, further confirming that the oscillations in the data recorded when the complete interferometry pulse sequence was applied arise from matter-wave interference of spatially separated Rydberg-atom de~Broglie waves. 

\section{Conclusion}\label{sec:conclusion}

In conclusion, we have presented, and realised experimentally, a scheme for matter-wave interferometry with atoms or molecules in high Rydberg states. For the pairs of Rydberg states employed in the experiments, with induced electric dipole moments of up to 11\,320~D and 11\,740~D, and hence differences in their electric dipole moments of up to 420~D, the difference in the forces exerted on the matter-wave components by the electric field gradients to which they were subjected were on the order of $10^{-24}$~N. The corresponding relative accelerations were $\sim100$~m/s$^2$. The typical displacements observed between the pairs of matter-wave components at the end of the interferometry sequences were up to 150~pm while the size of the atoms, given by $2\langle r \rangle$, where $\langle r \rangle$ is the expectation value the radial position operator acting on the Rydberg electron wavefunction, were 320~nm. From the range of experimental tests reported it is concluded that the maximal number of Rydberg-atom interference fringes observed was limited by the requirements for the adiabatic evolution of the Rydberg states in the combined electric and magnetic fields in which the experiments were performed. It is expected that a larger number of interference fringes could be seen if, for low values of $V_{\mathrm{grad}}$, a longer period of free flight is included between the pairs of electric field gradient pulses, or if lower-$\ell$ Rydberg states, for which the criteria for adiabatic evolution in the pulsed electric fields can be less restrictive, are employed. 

The Rydberg-atom interferometry scheme presented here relies on the generation of an inhomogeneous electric field distribution with a constant gradient in the direction of propagation of the atomic beam. If this gradient is accurately known, from the precise geometry of the electrode configuration, this type of interferometer represents a tool for measuring matter-wave de~Broglie wavelengths, and hence momenta. From this perspective, if forces are exerted on the atoms in the direction perpendicular to the surface of the Earth, it is expected that Rydberg-atom interferometry of the kind described here may be used to measure changes in de~Brogile wavelength caused by the acceleration of samples in Rydberg states in the gravitational field of the Earth. Such measurements are of interest at present for systems composed of antimatter, e.g., antihydrogen -- which is synthesised in high Rydberg states -- or positronium -- which in its ground state is too short lived ($142$~ns) to permit such measurements but when excited to Rydberg states exhibits significantly longer lifetimes ($>10~\mu$s) limited by spontaneous emission to the ground state. 

In addition to measurements of antimatter gravity, the large size ($320$~nm) of the Rydberg atoms used in the experiments reported here makes the methodology and results presented of interest for studies of effects of matter-wave interference in macroscopic quantum systems. In this context, future experiments directed toward investigations of the decoherence of the superpositions of Rydberg-atom momentum states with large spatial separations are of interest. 

\section{Acknowledgements}
	It is a pleasure to dedicate this article to Prof. Timothy P. Softley on the occasion of his 60th birthday. We are are very grateful to Prof. Softley for his support over the last years. We thank Dr. A. Deller (University College London) for valuable discussions about the experiments reported here. This work was supported by the Engineering and Physical Sciences Research Council under Grant No. EP/L019620/1, and the European Research Council (ERC) under the European Union's Horizon 2020 research and innovation program (Consolidator Grant No. 683341).


\begin{thebibliography}{99}

\bibitem{wing80a} W. H. Wing, Electrostatic Trapping of Neutral Atomic Particles, Phys. Rev. Lett. {\bf 45}, 631 (1980)
\bibitem{breeden81a} T. Breeden and H. Metcalf, Stark Acceleration of Rydberg Atoms in Inhomogeneous Electric Fields, Phys. Rev. Lett. {\bf 47}, 1726 (1981)
	
\bibitem{townsend01a}  D. Townsend, A. L. Goodgame, S. R. Procter, S. R. Mackenzie, and T. P. Softley, Deflection of krypton Rydberg atoms in the field of an electric dipole, J. Phys. B: At. Mol. Opt. Phys. {\bf 34}, 439 (2001)
	
\bibitem{gerlach22a} W. Gerlach and O. Stern, Der experimentelle Nachweis des magnetischen Moments des Silberatoms, Z. Phys. {\bf 8} 110 (1922)
	
\bibitem{gerlach22b} W. Gerlach and O. Stern, Der experimentelle Nachweis der Richtungsquantelung im Magnetfeld, Z. Phys. {\bf 9}, 349 (1922)
	
\bibitem{yamakita04a} Y. Yamakita, S. R. Procter, A. L. Goodgame, T. P. Softley, and F. Merkt, Deflection and deceleration of hydrogen Rydberg molecules in inhomogeneous electric fields, J. Chem. Phys. {\bf 121}, 1419 (2004)
	
\bibitem{allmendinger14a} P. Allmendinger, J. Deiglmayr, J. A. Agner, H. Schmutz, and F. Merkt, Surface-electrode decelerator and deflector for Rydberg atoms and molecules, Phys. Rev. A {\bf 90}, 043403 (2014)
	
\bibitem{lancuba13a} P. Lancuba and S. D. Hogan, Guiding Rydberg atoms above surface-based transmission lines, Phys. Rev. A {\bf 88}, 043427 (2013)
	
\bibitem{ko14a} H. Ko and S. D. Hogan, High-field-seeking Rydberg atoms orbiting a charged wire, Phys. Rev. A {\bf 89}, 053410 (2014)
	
\bibitem{deller16a} A. Deller, A. M. Alonso, B. S. Cooper, S. D. Hogan, and D. B. Cassidy, Electrostatically Guided Rydberg Positronium, Phys. Rev. Lett. {\bf 117}, 073202 (2016)
	
\bibitem{deller19a} A. Deller and S. D. Hogan, Confinement of high- and low-field-seeking Rydberg atoms using time-varying inhomogeneous electric fields, Phys. Rev. Lett. {\bf 122}, 053203 (2019)
	
\bibitem{alonso17a} A. M. Alonso, B. S. Cooper, A. Deller, L. Gurung, S. D. Hogan, and D. B. Cassidy, Velocity selection of Rydberg positronium using a curved electrostatic guide, Phys. Rev. A {\bf 95}, 053409 (2017)
	
\bibitem{vliegen06a} E. Vliegen, P. A. Limacher, and F. Merkt, Measurement of the three-dimensional velocity distribution of Stark-decelerated Rydberg atoms, Eur. Phys. J. D {\bf 40}, 73 (2006)
	
\bibitem{vliegen06b} E. Vliegen and F. Merkt, Normal-incidence electrostatic Rydberg atom mirror, Phys. Rev. Lett. {\bf 97}, 033002 (2006)
\bibitem{jones17a} A. C. L. Jones, J. Moxom, H. J. Rutbeck-Goldman, K. A. Osorno, G. G. Cecchini, M. Fuentes-Garcia, R. G. Greaves, D. J. Adams, H. W. K. Tom, A. P. Mills, Jr., and M. Leventhal, Focusing of a Rydberg Positronium Beam with an Ellipsoidal Electrostatic Mirror, Phys. Rev. Lett. {\bf 119}, 053201 (2017)
\bibitem{palmer17a} J. Palmer and S. D. Hogan, Experimental demonstration of a Rydberg-atom beam splitter, Phys. Rev. A {\bf 95}, 053413 (2017)
	
	
\bibitem{vliegen04a} E. Vliegen, H. J. W\"orner, T. P. Softley, and F. Merkt, Nonhydrogenic Effects in the Deceleration of Rydberg Atoms in Inhomogeneous Electric Fields, Phys. Rev. Lett. {\bf 92}, 033005 (2004)
\bibitem{vliegen05a} E. Vliegen and F. Merkt, On the electrostatic deceleration of argon atoms in high Rydberg states by time-dependent inhomogeneous electric fields, J. Phys. B: At. Mol. Opt. Phys. {\bf 38}, 1623 (2005)
	
	
\bibitem{vliegen07a} E. Vliegen, S. D. Hogan, H. Schmutz, and F. Merkt, Stark deceleration and trapping of hydrogen Rydberg atoms, Phys. Rev. A {\bf 76}, 023405 (2007)
\bibitem{hogan08a} S. D. Hogan and F. Merkt, Demonstration of Three-Dimensional Electrostatic Trapping of State-Selected Rydberg Atoms, Phys. Rev. Lett. {\bf 100}, 043001 (2008)
\bibitem{hogan09a} S. D. Hogan, Ch. Seiler, and F. Merkt, Rydberg-State-Enabled Deceleration and Trapping of Cold Molecules, Phys. Rev. Lett. {\bf 103}, 123001 (2009)
\bibitem{seiler11a} Ch. Seiler, S. D. Hogan, H. Schmutz, J. A. Agner, and F. Merkt, Collisional and Radiative Processes in Adiabatic Deceleration, Deflection, and Off-Axis Trapping of a Rydberg Atom Beam, Phys. Rev. Lett. {\bf 106}, 073003 (2011)
\bibitem{hogan12a} S. D. Hogan, P. Allmendinger, H. Sa\ss mannshausen, H. Schmutz, and F. Merkt, Surface-Electrode Rydberg-Stark Decelerator, Phys. Rev. Lett. {\bf 108} 063008 (2012)
\bibitem{lancuba16a}  P. Lancuba and S. D. Hogan, Electrostatic trapping and in situ detection of Rydberg atoms above chip-based transmission lines, J. Phys. B: At. Mol. Opt. Phys. {\bf 49} 074006 (2016)
	
\bibitem{allmendinger16a} P. Allmendinger, J. Deiglmayr, O. Schullian, K. H\"oveler, J. A. Agner, H. Schmutz, and F. Merkt, New method to study ion-molecule reactions at low temperatures and application to the $\mathrm{H}_2^+ + \mathrm{H}_2\rightarrow \mathrm{H}_3^+ + \mathrm{H}$ reaction, Chem. Phys. Chem. {\bf 17} 3596 (2016)
\bibitem{allmendinger16b} P. Allmendinger, J. Deiglmayr, K. H\"oveler, O. Schullian, and F. Merkt, Observation of enhanced rate coefficients in the $\mathrm{H}_2^+ + \mathrm{H}_2\rightarrow \mathrm{H}_3^+ + \mathrm{H}$ reaction at low collision energies, J. Chem. Phys. {\bf 145} 244316 (2016)
	
\bibitem{seiler16a}  Ch. Seiler, J. A. Agner, P. Pillet, and F Merkt, Radiative and collisional processes in translationally cold samples of hydrogen Rydberg atoms studied in an electrostatic trap, J. Phys. B: At. Mol. Opt. Phys. {\bf 49} 094006 (2016)
	
\bibitem{cassidy18a} D. B. Cassidy, Experimental progress in positronium laser physics, Eur. Phys. J. D {\bf 72}, 53 (2018)	
	
\bibitem{bassi13a} A. Bassi, K. Lochan, S. Satin, T. P. Singh, and H. Ulbricht, Models of wave-function collapse, underlying theories, and experimental tests, Rev. Mod. Phys. {\bf 85}, 471 (2013)
		
\bibitem{mills02a} A. P. Mills, Jr. and M. Leventhal, Can we measure the gravitational free fall of cold Rydberg state positronium?, Nucl. Instrum. Methods Phys. Res., Sect. B {\bf 192}, 102 (2002)
	
\bibitem{kellerbauer08a}A. Kellerbauer, M. Amoretti, A. S. Belov, G. Bonomi, I. Boscolo, R. S. Brusa, M. B\"uchner, V. M. Byakov, L. Cabaret, C. Canali, C. Carraro, F. Castelli, S. Cialdi, M. de Combarieu, D. Comparat, G. Consolati, N. Djourelov, M. Doser, G. Drobychev, A. Dupasquier, G. Ferrari, P. Forget, L. Formaro, A. Gervasini, M. G. Giammarchi, S. N. Gninenko, G. Gribakin, S. D. Hogan, M. Jacquey, V. Lagomarsino, G. Manuzio, S. Mariazzi, V. A. Matveev, J. O. Meier, F. Merkt, P. Nedelec, M. K. Oberthaler, P. Pari, M. Prevedelli, F. Quasso, A. Rotondi, D. Sillou, S. V. Stepanov, H. H. Stroke, G. Testera, G. M. Tino, G. Tr\`enec, A. Vairo, J. Vigu\`e, H. Walters, U. Warring, S. Zavatarelli, D. S. Zvezhinskij, Proposed antimatter gravity measurement with an antihydrogen beam, Nucl. Instrum. Methods Phys. Res., Sect. B {\bf 266}, 351 (2008)
	
	
\bibitem{amole13a} C. Amole, M. D. Ashkezari, M. Baquero-Ruiz, W. Bertsche, E. Butler, A. Capra, C. L. Cesar, M. Charlton, S. Eriksson, J. Fajans, T. Friesen, M. C. Fujiwara, D. R. Gill, A. Gutierrez, J. S. Hangst, W. N. Hardy, M. E. Hayden, C. A. Isaac, S. Jonsell, L. Kurchaninov, A. Little, N. Madsen, J. T. K. McKenna, S. Menary, S. C. Napoli, P. Nolan, A. Olin, P. Pusa, C. \O Rasmussen, F. Robicheaux, E. Sarid, D. M. Silveira, C. So, R. I. Thompson, D. P. van der Werf, J. S. Wurtele, and A. I. Zhmoginov, Description and first application of a new technique to measure the gravitational mass of antihydrogen, Nat. Commun. {\bf 4}, 1785 (2013)
	
\bibitem{cassidy14a} D. B. Cassidy and S. D. Hogan, Atom control and gravity measurements using Rydberg positronium, Int. J. Mod. Phys. Conf. Ser. {\bf 30}, 1460259 (2014)
	
\bibitem{cronin09a} A. D. Cronin, J. Schmiedmayer, and D. E. Pritchard, Optics and interferometry with atoms and molecules, Rev. Mod. Phys. {\bf 81}, 1051 (2009)
	
\bibitem{hornberger12a} K. Hornberger, S. Gerlich, P. Haslinger, S. Nimmrichter, and M. Arndt, M., Colloquium: Quantum interference of clusters and molecules, Rev. Mod. Phys. {\bf 84}, 157 (2012)
	
	
\bibitem{bohm51a} D. Bohm, \emph{Quantum Theory}, Prentice-Hall, New York (1951)
\bibitem{wigner63a} E. P. Wigner, The Problem of Measurement, Am. J. Phys. {\bf 31}, 6 (1963)
\bibitem{schwinger88a} J. Schwinger, M. O. Scully, and B.-G. Englert, Is spin coherence like Humpty-Dumpty?, Z Phys. D {\bf 10}, 135 (1988)
	
\bibitem{sokolov70a} Y. L. Sokolov, Influence of Lamb shift on the interference of excited states of hydrogen atoms,  JETP Lett. {\bf 11}, 359 (1970)
	
\bibitem{sokolov73a} Y. L. Sokolov, Interference of the 2p$_{1/2}$ state of the hydrogen atom, Sov. Phys. JETP {\bf 36}, 243 (1973)
	
\bibitem{miniatura91a} Ch. Miniatura, F. Perales, G. Vassilev, J. Reinhardt, J. Robert and J. Baudon, A longitudinal Stern-Gerlach interferometer : the ``beaded'' atom, J. Phys. II France {\bf 1} 425 (1991) 
	
\bibitem{miniatura92a} Ch. Miniatura, J. Robert, O. Gorceix, V. Lorent, S. Le Boiteux, J. Reinhardt, and J. Baudon, Atomic interferences and the topological phase, Phys. Rev. Lett. {\bf 69}, 261 (1992)
	
\bibitem{nicChormaic93a} S Nic Chormaic, V Wiedemann, C Miniatura, J Robert, S Le Boiteux, V Lorent, O Gorceix, S Feron, J Reinhardt and J Baudon, Longitudinal Stern-Gerlach atomic interferometry using velocity selected atomic beams, J. Phys. B {\bf 26}, 1271 (1993)
	
\bibitem{machluf13a} S. Machluf, Y. Japha, and R. Folman, Coherent Stern-Gerlach momentum splitting on an atom chip, Nat. Commun. {\bf 4}, 2424 (2013)
	
	
\bibitem{halfmann00a} T. Halfmann, J. Koensgen, and K. Bergmann, A source for a high-intensity pulsed beam of metastable helium atoms, Meas. Sci. Technol. {\bf{11}}, 1510 (2000).
	
\bibitem{delande88a} D. Delande and J. C. Gay, A New Method for Producing Circular Rydberg States, Euro. Phys. Lett. {\bf 5}, 303 (1988)
	
\bibitem{zhelyazkova16a} V. Zhelyazkova and S. D. Hogan, Preparation of circular Rydberg states in helium using the crossed-fields method, Phys. Rev. A {\bf 94}, 023415 (2016)
	
\bibitem{hogan18a} S. D. Hogan, Y. Houston, and B. Wei, Laser photoexcitation of Rydberg states in helium with $n>400$, J. Phys. B {\bf 51}, 145002 (2018)
	
\bibitem{pauli26a} W. Pauli, \"Uber das Wasserstoffspektrum vom Standpunkt der neuen Quantenmechanik, Z. Phys. {\bf 36}, 336 (1926)
	
\bibitem{elliott95a} R. J. Elliott, G. Droungas, and J.-P. Connerade, Active cancellation of the motional Stark effect in the diamagnetic spectrum of Ba, J. Phys. B {\bf 28}, L537 (1995)

\bibitem{zhelyazkova15b} V. Zhelyazkova and S. D. Hogan, Rydberg-Stark states in oscillating electric fields, Mol. Phys. {\bf 113}, 3979 (2015) 

\bibitem{zhelyazkova15a} V. Zhelyazkova and S. D. Hogan, Probing interactions between Rydberg atoms with large electric dipole moments in amplitude-modulated electric fields, Phys. Rev. A {\bf 92}, 011402(R) (2015)

\end{thebibliography}
\end{document}